\documentclass[aip,JAP,showpacs,reprint,floatfix,superscriptaddress]{revtex4-1}
\pdfoutput=1
\usepackage[utf8]{inputenc}
\usepackage{graphicx}
\usepackage{dcolumn}
\usepackage{amsmath,amssymb,bm}
\usepackage{esint}
\usepackage[caption=false]{subfig}
\usepackage{color}
\usepackage[]{lineno}

\usepackage[dvipsnames,svgnames,x11names]{xcolor}
\usepackage[final]{changes}
\definechangesauthor[color=Red]{GW}

\begin{document}
\title{
Controlling $T_c$ of Iridium Films Using the Proximity Effect
}

\author{R. Hennings-Yeomans}
\email{hennings@berkeley.edu}
\affiliation{Department of Physics, University of California, Berkeley, CA 94720 USA}
\affiliation{Nuclear Science Division, Lawrence Berkeley National Laboratory, Berkeley, CA 94720 USA}

\author{C.L. Chang}
\affiliation{High Energy Physics Division, Argonne National Laboratory, Argonne, IL 60439 USA}
\affiliation{Kavli Institute for Cosmological Physics, University of Chicago, Chicago, IL 60637 USA}
\affiliation{Department of Astronomy and Astrophysics, University of Chicago, Chicago, IL 60637 USA}

\author{J. Ding}
\affiliation{Materials Science Division,
Argonne National Laboratory, Argonne, IL 60439 USA}

\author{A. Drobizhev}
\affiliation{Department of Physics, University of California, Berkeley, CA 94720 USA}
\affiliation{Nuclear Science Division, Lawrence Berkeley National
Laboratory, Berkeley, CA 94720 USA}

\author{B.K. Fujikawa}
\affiliation{Nuclear Science Division, Lawrence Berkeley National Laboratory, Berkeley, CA 94720 USA}

\author{S. Han}
\affiliation{Department of Physics, University of California, Berkeley, CA 94720 USA}

\author{G. Karapetrov}
\affiliation{Department of Physics, Drexel University, Philadelphia, PA 19104 USA}

\author{Yu.G. Kolomensky}
\affiliation{Department of Physics, University of California, Berkeley, CA 94720 USA}
\affiliation{Nuclear Science Division, Lawrence Berkeley National Laboratory, Berkeley, CA 94720 USA}

\author{V. Novosad}
\affiliation{Materials Science Division, Argonne National Laboratory, Argonne, IL 60439 USA}
\affiliation{Physics Division, Argonne National Laboratory, Argonne, IL 60439}

\author{T. O'Donnell}
\affiliation{Department of Physics, University of California, Berkeley, CA 94720 USA}
\affiliation{Nuclear Science Division, Lawrence Berkeley National Laboratory, Berkeley, CA 94720 USA}

\author{J.L. Ouellet}
\affiliation{Department of Physics, University of California, Berkeley, CA 94720 USA}
\affiliation{Massachusetts Institute of Technology, Cambridge, MA 02139 USA}

\author{J. Pearson}
\affiliation{Materials Science Division,
Argonne National Laboratory, Argonne, IL 60439 USA}

\author{T. Polakovic}
\affiliation{Department of Physics, Drexel University, Philadelphia, PA 19104 USA}
\affiliation{Physics Division, Argonne National Laboratory, Argonne, IL 60439}

\author{D. Reggio}
\affiliation{Department of Physics, University of California, Berkeley, CA 94720 USA}

\author{B. Schmidt}
\affiliation{Nuclear Science Division, Lawrence Berkeley National Laboratory, Berkeley, CA 94720 USA}

\author{B. Sheff}
\affiliation{Department of Physics, University of California, Berkeley, CA 94720 USA}

\author{V. Singh}
\affiliation{Department of Physics, University of California, Berkeley, CA 94720 USA}

\author{R.J. Smith}
\affiliation{Department of Physics, University of California, Berkeley, CA 94720 USA}

\author{G. Wang}
\email{gwang@anl.gov}
\affiliation{High Energy Physics Division, Argonne National Laboratory, Argonne, IL 60439 USA}

\author{B. Welliver}
\affiliation{Nuclear Science Division, Lawrence Berkeley National
Laboratory, Berkeley, CA 94720 USA}

\author{V.G. Yefremenko}
\affiliation{High Energy Physics Division, Argonne National Laboratory, Argonne, IL 60439 USA}

\author{J. Zhang}
\affiliation{High Energy Physics Division, Argonne National Laboratory, Argonne, IL 60439 USA}

\date{\today}

\begin{abstract}
A superconducting Transition-Edge Sensor (TES) with low-$T_c$ is essential in a high resolution calorimetric detection.
With a motivation of developing sensitive calorimeters for applications in cryogenic neutrinoless double beta decay searches, we have been investigating methods to reduce the $T_c$ of an Ir film down to 20 mK.
Utilizing the proximity effect between a superconductor and a normal metal, we found two room temperature fabrication recipes of making Ir-based low-$T_c$ films.
In the first approach, an Ir film sandwiched between two Au films, a Au/Ir/Au trilayer, has a tunable $T_c$ in the range of 20-100~mK depending on the relative thicknesses.
In the second approach, a paramagnetic Pt thin film is used to create Ir/Pt bilayer with a tunable $T_c$ in the same range.
We present detailed study of fabrication and characterization of Ir-based low-$T_c$ films, and compare the experimental results to theoretical models.
We show that Ir-based films with predictable and reproducible critical temperature can be consistently fabricated for use in large scale detector applications.
\end{abstract}


\maketitle

\section{Introduction} \label{Sec:Intro}
Neutrinoless Double Beta Decay (NLDBD) searches~\cite{Giuliani:12, Cremonesi:14} explore fundamental physics beyond the standard model.
With two leptons (electrons) in the final state and none in the initial, observation of NLDBD would imply lepton flavor violation by two units and provide strong constraints for theories beyond the Standard Model of particle physics~\cite{Third, Fourth, Fifth, Sixth}.
This would constitute a direct proof that neutrinos are their own anti-particles, and therefore Majorana fermions~\cite{Sixth, Second}.
In CUORE experiment~\cite{Alduino:18}, a ton-scale cryogenic calorimetric array is currently searching for NLDBD decay.
The dominant background in CUORE arises from so-called degraded $\alpha$ particles~\cite{cuore_background_model}.
To identify and eliminate this background, the next generation experiment called CUPID~\cite{CUPID} aims to discriminate this background through the use of particle ID.
CUPID will use scintillating Li$_2$MoO$_4$ crystals~\cite{Bekker16} as the primary calorimeters with light detectors instrumented nearby. Since $\alpha$ events produce less scintillation light than $\beta$ and $\gamma$ events, a measure of both heat and scintillation light can provide a powerful discrimination tool to reject $\alpha$ events.
In the CUPID baseline design, discriminating a NLDBD event from backgrounds requires a detector that has both good energy resolution\added[id=GW]{ ($\sigma$$\sim$20-30 eV)} and fast response time\added[id=GW]{ (as fast as $\tau$$\sim$10 $\mu$s)~\cite{cupidpcdr}}.
A TES calorimetric detector could potentially be used in the CUPID experiment for measuring light (and heat), as TES-based experiments have been shown to have both excellent energy resolution and fast timing parameters~\cite{KentIrwin, IrwinHilton, Ullom:15, CRESST:16}.

There are three components needed to construct a TES calorimeter to measure scintillation light: a surface-engineered \added[id=GW]{large area }silicon\added[id=GW]{ or germanium} wafer to act as a light absorber, a TES thermometer to measure the absorbed energy, and a weak thermal link to a cold bath. For an energy $E$ deposited on the TES detector, the temperature change of the TES is $\Delta T = E/C$, where $C$ is the detector's total\deleted[id=GW]{ specific} heat\added[id=GW]{ capacity}. Operated under negative electro-thermal feedback~\cite{KentIrwin}, a TES calorimeter has an expected energy resolution of
\begin{equation}
\Delta E \approx 2.35 \sqrt{\frac{4k_B T_c^2 C}{\alpha} \sqrt{\frac{n}{2}}},
\label{eq:resolution}
\end{equation}
where $k_B$ is the Boltzmann constant, $T_c$ is the TES transition temperature, and $n$ is an index varying between three and five depending on the thermal carriers in the weak link~\cite{IrwinHilton, Giazotto:06}. Here $\alpha \approx (T_c/R)(dR/dT)$ is a parameter characterizing the TES superconducting-to-resistive transition profile, where $R$ is the temperature-dependent resistance of the TES in the transition.

At low temperatures the total\deleted[id=GW]{ specific} heat\added[id=GW]{ capacity} of the calorimeter consists of the\deleted[id=GW]{ specific} heat\added[id=GW]{ capacity} of the absorber (silicon wafer) dominated by its phonon contribution ($C_{ph}(T)\propto T^3$) and the\deleted[id=GW]{ specific} heat\added[id=GW]{ capacity} of the thin film TES thermometer dominated by the electronic contribution ($C_{el}(T)\propto T$). Although the electronic specific heat is dominant at T$<$~1\,K, in our case with large area silicon light detector the mass of the silicon wafer is several orders of magnitude larger than the mass of the thin film thermometer. Therefore, the temperature dependence of the energy resolution in equation~\ref{eq:resolution} will be dominated by the $C_{ph}(T)$ of the phonons in the silicon wafer, resulting in $\Delta E \propto T^{5/2}$.

The second key parameter of a TES calorimeter is the effective time constant~\cite{IrwinHilton},
\begin{equation}
\tau_{eff} \approx \frac{C}{G}\frac{1}{1+\mathcal{L}} \ ,
\label{eq:taue}
\end{equation}
where C/G is the natural thermal time constant due to the calorimeter's \deleted[id=GW]{specific }heat\added[id=GW]{ capacity} $C$ and TES's thermal coupling to the cold bath characterized by the thermal conductance $G$.
$\mathcal{L} \approx \alpha P_J / G T_c$ is the thermal loop gain of the TES operated under electro-thermal feedback~\cite{IrwinHilton} with TES Joule heating $P_J$.
The TES detector time constant is reduced by the thermal loop gain $\mathcal{L}$ (typically between 1 and 100) which depends on the superconducting transition temperature $T_c$ and the transition profile.

According to equations~\ref{eq:resolution} and~\ref{eq:taue}, a calorimeter suitable for NLDBD favors a TES with a low-$T_c$ and a large $\alpha$.
\replaced[id=GW]{Common}{Low-$T_c$} TES materials \replaced[id=GW]{such as}{obtained by sputtering include W films~\cite{Roth08},} dilute AlMn alloy films~\cite{Deiker04}, and various bilayers such as Ti/Au~\cite{Yoshino08}, Mo/Cu~\cite{Ullom04}, and Mo/Au~\cite{Smith08}\deleted[id=GW]{. However, none of these TES films} have \added[id=GW]{not} demonstrated sufficiently low $T_c$ ($<$ 30\,mK) with good reproducibility for applications in a cryogenic NLDBD search experiment. \added[id=GW]{TES made from E-beam evaporated W shows promise, as demonstrated by the CRESST collaboration where a W TES with 15 mK $T_c$ was used with a 23.6 g CaWO$_4$ detector to realize an energy threshold as low as 38.1 eV~\cite{abdelhameed2019}. However, reliable tuning of the W $T_c$ below 50 mK is an active area of research as there are both alpha and beta phases~\cite{Roth08, lita2005} in sputtered W films. Furthermore, there has not yet been a demonstration of the scalability of a TES process to thousands of devices for applications in such low temperature.}

In this paper, we present an alternative approach using a thin iridium film with an intrinsic low $T_c$ and covering it with normal metals to further reduce its $T_c$ through the proximity effect~\cite{Usadel70,Wang17,Martinis,Kupriyanov88,Golubov1,Radovic,Brammertz,Golubov2,Khusainov,Fominov,Zhao18}.
Iridium is a good low-$T_c$ candidate material because of its excellent chemical stability, low bulk superconducting transition temperature of 110-140\,mK~\cite{Hein62,Galeazzi09} and large coherence length ($\xi(0)=$4.4\,$\mu$m~\cite{Gubser73}).
The $T_c$ of an Ir thin film can be further reduced by putting it in proximity to a normal metal, such as gold.
Tuning the relative layer thicknesses of the gold and iridium films allows for adjustment of the bilayer $T_c$ to a specific desired value. \added[id=GW]{Prior work using E-beam evaporated Ir bilayers has achieved $T_c$ as low as 33 mK ~\cite{CRESST-films-2, CRESST-films-1, CRESST-films-3}, but these films required substrate heating, which limits the applicability of these films to certain substrates. Importantly, this paper suggests that E-beam film growth may be susceptible to impurity contamination. Other techniques for Ir film growth include pulse laser deposition\cite{galeazzi2004} and RF sputtering deposition~\cite{fukuda2017, bogorin2008}. Most of these techniques also require substrate heating and none of them have demonstrated $T_c$ below 50 mK.}
\deleted[id=GW]{Previous work on Ir/Au bilayers suggests that to reach a sufficiently low-$T_c$, the substrate should be heated to a temperature between 500 and  600$^{\circ}$C during Ir film deposition~\cite{CRESST-films-2, CRESST-films-1, CRESST-films-3}.}

We present results from our $T_c$ studies of Ir-based bilayers and trilayers.
Our studies demonstrate that Ir-based proximity films can have a desirable low $T_c$ using sputtering deposition at room temperature.
The paper is structured as follows: in section~\ref{Sec:FabAndSetup}, we describe the fabrication of films and the $T_c$ measurement method.
In section~\ref{Sec:Discussion}, we discuss the data obtained on Ir/Pt bilayers and Au/Ir/Au trilayers grown on substrates at room temperature, and compare the data on Ir/Au and Ir/Pt bilayers with Ir films grown at elevated substrate temperatures.
\section{Fabrication and Characterization} \label{Sec:FabAndSetup}
\subsection{Thin film fabrication}
The thin film samples presented in this paper were fabricated on two-inch high-resistivity ($>$10000 $\Omega$-cm) silicon wafers using DC magnetron sputtering at the Materials Sciences Division at Argonne National Laboratory.
The samples were made using an AJA 2400 sputtering system\replaced[id=GW]{, which contains}{containing} five two-inch sources in the main sputtering chamber\added[id=GW]{, is configured for direct deposition with normal incidence orientation, and uses a cryopump for a high vacuum}.
In our experiments we used 4N Ir target, 4N Pt target, and a 4N Au target.
The base pressure in the sputtering chamber was lower than $1 \times 10^{-7}$~mbar.
The sputtering was performed with Ar gas at 3\,mbar. All depositions were done with the same Ar pressure and sputtering power. Deposition rates of 2.6~\AA/sec for Ir, 2.9~\AA/sec for Au and 2.1~\AA/sec for Pt films were used.
We fabricated bilayers of Ir/Pt and Ir/Au along with trilayers of Au/Ir/Au with varying layer thicknesses.
Most of the samples were grown at room temperature with a fraction of bilayer samples grown at an elevated substrate temperature during the Ir film deposition.
For the Au/Ir/Au trilayer samples, a 3~nm thick iridium layer was deposited prior to the trilayer to promote adhesion to the silicon wafer.
For all the $T_c$ samples, the thickness of each film have an uncertainty of less than 3\%.

After all films were deposited, the wafers were diced into squares of 3 mm per side. The chips were attached to a copper plate using GE-varnish and wire bonded in 4-wire measurement configuration for superconducting-to-resistive transition measurements.
\subsection{Electrical transport characterization}
The film samples were installed on the mixing chamber plate of an Oxford Triton~400 dilution refrigerator unit at the Physics Department of the University of California at Berkeley.
A 4-wire resistance measurement was performed using a Lakeshore model 370 AC resistance bridge.
A 13.7~Hz excitation current was injected into one pair of leads while the voltage was measured on the other pair using
lock-in technique.
Relative uncertainty of the resistance measurement was between 0.05-0.1$\%$.

Temperature measurements in the range of 50-200\,mK used a calibrated ruthenium oxide (RuO$_2$) thermometer from Lakeshore cryotronics.
In the temperature range 30-50\,mK this thermometer was calibrated against a nuclear demagnetization $^{60}$Co thermometer mounted at the center of the mixing chamber plate.
Between 8-30\,mK, we utilized a Johnson noise thermometer from Magnicon~\cite{NT-magnicon} that was calibrated against the $^{60}$Co thermometer.
Systematic uncertainty of the temperature measurement in 30-200~mK range was less than 1$\%$ and in the range between 8-30\,mK was less than 0.5\,mK.

\begin{figure}[hbt!]
	\includegraphics[width = 3 in]{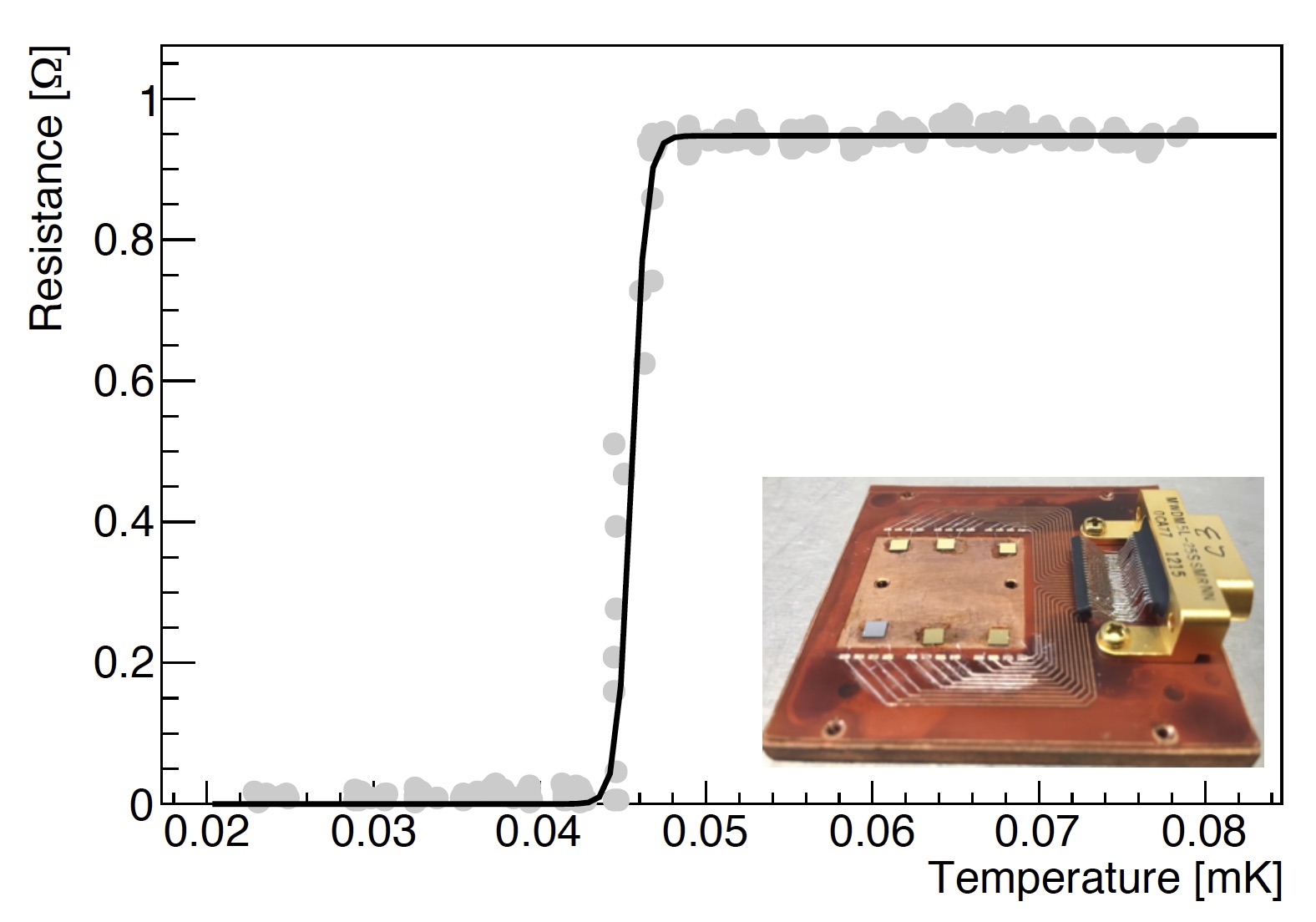}
	\caption{Resistance vs temperature of an Ir/Pt bilayer (100~nm Ir and 60~nm Pt).
	The dots are experimental data from scanning up and down in temperature.
	The solid line is a fit to the data using the empirical equation~\ref{eq:Tc_fit_function}.
	This bilayer has a critical temperature T$_c$=45.5$\pm$0.7\,mK at 50\% of $R_n$.
	The inset shows a photograph of the cryogenic mount for six superconducting samples, each bonded for a four-wire measurement of resistance.}
	\label{fig:tc_fit}
\end{figure}

The critical temperature of the superconducting films was determined by a least squares fit of the measured resistance vs temperature data to an empirical equation
\begin{eqnarray}
R(T)=\frac{R_n}{1+e^{\left( AT+B\right)}} + C ,
\label{eq:Tc_fit_function}
\end{eqnarray}
where $T$ is the mixing chamber plate temperature, $R$ is the measured sample resistance as a function of $T$, $R_n$ is the normal resistance just above $T_c$, and C can be a nonzero parasitic resistance.
The critical temperature is evaluated at $R$ = 50\% of $R_n$, which is T$_c$ = -B/A.
Fig.~\ref{fig:tc_fit} shows the comparison between the measured resistance vs temperature $R(T)$ of an Ir/Pt (100\,nm/60\,nm) bilayer and the fit using equation~\ref{eq:Tc_fit_function}.

The $T_c$ measurements were made using two excitation currents, 3.16~$\mu$A for films with higher resistances (Ir bare films, Ir/Au and Ir/Pt bilayers) and 31.6~$\mu$A for films with lower resistances (Au/Ir/Au trilayers).
The dominant systematic error for the $T_c$ measurement is the difference in temperature between the mixing chamber plate and the superconducting sample arising primarily from the electron-phonon decoupling resistance~\cite{Giazotto:06} of the film, which is discussed further in section~\ref{electron_phonon}.

\begin{figure}[hbt!]
\begin{center}
\includegraphics[width=3 in]{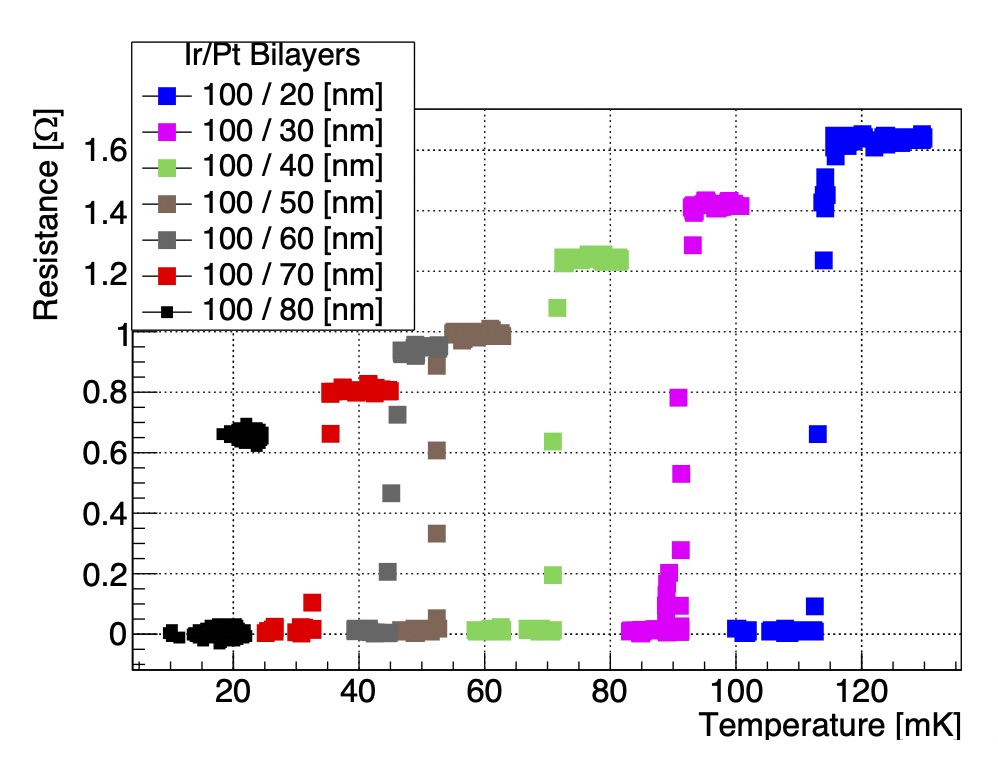}
\includegraphics[width=3 in]{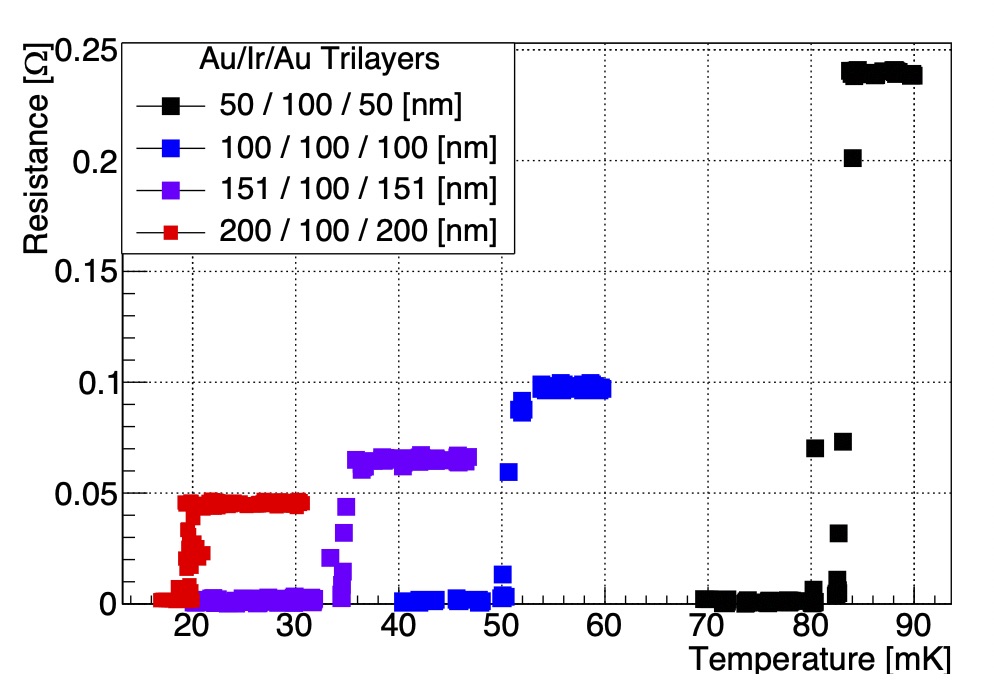}
\caption{
Top: Measured resistance vs temperature for Ir/Pt bilayers.
The thickness of the Ir layer was kept constant at 100\,nm while the Pt layer was varied between 20-80\,nm. The excitation current was 3.16\,$\mu$A.
Bottom: Measured resistance vs temperature for Au/Ir/Au trilayers.
The thickness of the Ir layer was kept constant at 100~nm while the thicknesses of both Au layers was varied between 50-200\,nm. The excitation current was 31.6~$\mu$A. \added[id=GW]{Note that the measured normal resistance of each sample approximately equals to its sheet resistance.}
}
\label{fig:two_tc}
\end{center}
\end{figure}

Additional electrical transport measurements were carried out at Argonne National Laboratory with a Bluefors LD-400 dilution refrigerator.
A RuO$_2$ thermometer calibrated down to 7\,mK was used for temperature measurement.
Lakeshore 372 AC bridge with excitation of 0.3 $\mu$A was used for the resistance measurement.

We have measured superconducting critical temperatures $T_c$ of Ir/Pt bilayers and Au/Ir/Au trilayers sputtered at room temperature (Fig.~\ref{fig:two_tc}).
Every sample has the same Ir thickness (100~nm) while the thicknesses of the normal metals were varied.
For each trilayer, the top and bottom Au film have the same thickness.
\section{Discussion}  \label{Sec:Discussion}
Analysis of our data provides estimates for underlying thin film material properties including the electron-phonon thermal resistance, electron interface transparency and electron spin flip relaxation time.
By comparing our data with additional measurements of films grown by different fabrication processes, we can also investigate the stability of our thin film growth process.
\subsection{Electron-phonon thermal impedance in Ir/Pt bilayers} \label{electron_phonon}
From our electrical transport data we could estimate the electron-phonon decoupling thermal resistance~\cite{Giazotto:06} in Ir/Pt films at these low temperatures.
We performed $T_c$ measurements at different excitation currents for the Ir/Pt bilayer (100\,nm/80\,nm) and observed a dependence of the critical temperature on the excitation current.
Decreasing the bias current from 3.16~$\mu$A to 0.316~$\mu$A shifted $T_c$ from 20.9$\pm$0.03~mK to 23.0$\pm$0.05~mK.
\added[id=GW]{The measured $T_c$ error due to stray magnetic field of the excitation current is negligibly small.}
For electron-phonon decoupling, the Joule heating power is given by
\begin{equation}
P_J = \Sigma V (T_c^5-T_b^5),
\label{eq:heat}
\end{equation}
where $\Sigma$ is a constant characterizing the thermal resistance between electrons and phonons, $V$ is the sample volume, and $T_b$ is the bath temperature.
\replaced[id=GW]{From}{Using} the measured temperature difference between 3.16~$\mu$A and 0.316~$\mu$A excitations, we \replaced[id=GW]{can use equation~\ref{eq:heat} to infer}{measure} $\Sigma \approx 0.88 \times 10^9\ W/m^3K^5$\deleted[id=GW]{ in our system}, which is \replaced[id=GW]{a reasonable value for}{in the range of} electron-phonon decoupling in metal films~\cite{Giazotto:06}.

\subsection{Interface transparency and spin flip time in Pt}
\begin{figure}[hbt!]
\begin{center}
\includegraphics[width=3.2 in]{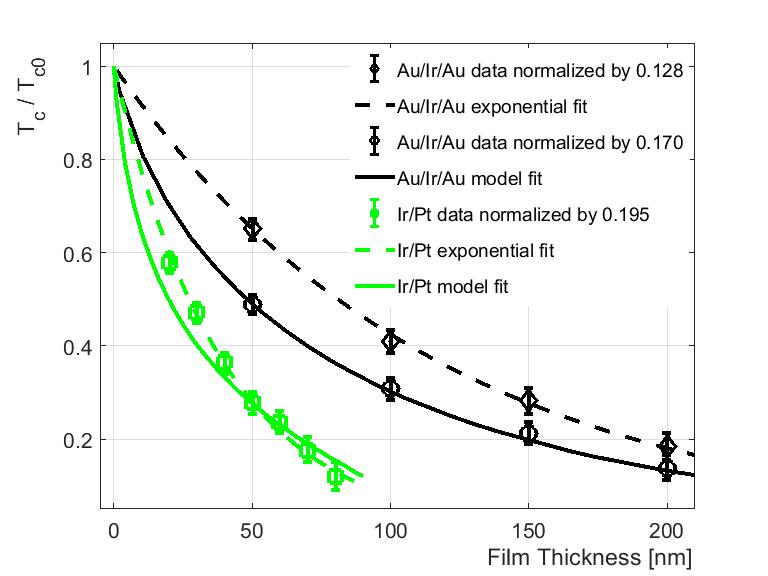}
\caption{
Reduced transition temperature ($T_c / T_{c0}$) vs thickness of Au in Au/Ir/Au trilayers (black symbols) and thickness of Pt in Ir/Pt bilayers (green symbols).
In all data the thicknesses of Ir film is kept at 100\,nm.
The $T_c$ of Au/Ir/Au has small Joule heating power correction incorporated using equation~\ref{eq:heat} with $\Sigma = 2.4 \times 10^9\ W/m^3K^5$ for Au~\cite{Giazotto:06}.
The $T_c$ of Ir/Pt has a small Joule heating correction accounted for by using the electron-phonon decoupling strength in section~\ref{electron_phonon}.
The error bars include the $T_{c}$ uncertainties from measurements and the $T_{c}$ uncertainties interpolated from the 3\% film thickness uncertainties.
The black line is for a model~\cite{Wang17} with a $T_{c0}=170$~mK and an interface electron transparency $t=0.105$.
Black diamonds are for $T_c / T_{c0}$ with $T_{c0}$=128~mK from an exponential fit resulting in a $\chi^2$=0.37 with 2 degrees of freedom.
The dashed black line is the exponential function, $T_c=T_{c0}e^{-d/d_0}$, where $d$ is Au thickness and $d_0$=116.8~nm.
The green line is for a model~\cite{Wang17} with a measured $T_c=195$~mK of 100~nm Ir film on silicon, the same interface electron transparency for the Au/Ir/Au trilayers, and an electron spin flip time $\tau_s=0.116~ns$.
The dashed green line is a exponential function, $T_c=T_{c0}e^{-d/d_0}$, where $d$ is Pt thickness and $d_0$=39.3~nm.
This exponential fit results in a $\chi^2$=1.95 with 6 degrees of freedom.
}
\label{fig:reducedTc_vs_thickness}
\end{center}
\vspace{-10pt}
\end{figure}

The measured $T_c$ data of Au/Ir/Au trilayers and Ir/Pt bilayers in Fig.~\ref{fig:reducedTc_vs_thickness} are well described by a single exponential relation to the thicknesses of normal metal films.
By fitting our data to a more complex proximity model~\cite{Wang17}, we can interpret our measurements to estimate underlying materials properties in our structures at these low temperatures, including the interface transparency and the spin flip time in Pt.
However, since a single exponential model is already a good fit, the results of our more complex modeling for Ir/Pt bilayers are not well constrained.

For this analysis, we use a proximity model~\cite{Wang17}, which allows analytical $T_c$ calculation of an Ir-based bilayer or trilayer and was developed with proximity theory~\cite{Usadel70,Martinis,Kupriyanov88}.
With transition temperature $T_{c0}$ of bare Ir film, electron density of states, thicknesses of Ir and normal metal films, electron transparency across superconductor/normal metal interface, and electron spin flip time in the case of Pt, the model calculates the transition temperature $T_c$ of a bilayer or trilayer.
Fig. 3 shows the reduced transition temperature $T_c / T_{c0}$ dependence on the normal metal thickness along with predictions from our model.

In the model, the strong $T_c$ suppression of the Pt film comes from two effects.
The first one is from the electron density of states at Fermi level~\cite{John84},
\begin{eqnarray}
n_{N,S} = 3 \gamma_{N,S} / \pi^2 k_B^2 (1+\lambda_{N,S}),
\label{eq:density}
\end{eqnarray}
where $\gamma$ is the electronic specific heat coefficient of a metal, $\lambda$ is electron phonon coupling constant, $N$ and $S$ are for normal metal and superconductor respectively.
In the model, the electron density of states of Pt ($7.27 \times 10^{47} / J \cdot m^{3}$) is much larger than that of Au ($9.43 \times 10^{46} / J \cdot m^{3}$).
The second effect comes from paramagnetic properties of Pt~\cite{Katayama03, Herrmannsd96, Herrmannsd962, Entin75, hauser66, Andres68, Jensen68}, which can be characterized by an effective electron spin flip time.

To calculate the $T_c$ of Au/Ir/Au trilayers, equations 10, 11 and 13 from Wang et al.~\cite{Wang17} together with equation~\ref{eq:density} in this paper were utilized.
The trilayer model has two independent parameters: the $T_{c0}$ of the bare Ir film and the electron transparency, $t$, at the interface between the Ir and Au films.
We assume that the electron transparency of the top Ir/Au interface is identical to the transparency of the bottom Ir/Au interface.
The black curve in Fig.~\ref{fig:reducedTc_vs_thickness} corresponds to $T_{c0} \approx $ 170 mK and electron transparency $t \approx $ 0.105.
This curve is a reasonable fit to the data having a $\chi^2$ between the data and the model of 0.46 for 2 degrees of freedom.

In the calculations of $T_c$ of Ir/Pt bilayers, equations 13 - 16 from Wang et al.~\cite{Wang17} together with equation~\ref{eq:density} were utilized.
The bilayer model has two independent parameters: the barrier transparency, $t$, at the interface between the Ir and Pt films and the effective electron spin flip time, $\tau_s$.
Larger electron interface transparency and faster electron spin flip rate similarly lead to more $T_c$ suppression by the normal metal.
We assume that the Ir/Pt interface transparency is identical to that of Ir/Au for removing any possible degeneracy between $t$ and $\tau_s$.
The green curve in Fig.~\ref{fig:reducedTc_vs_thickness} has a spin flip time of $\tau_s$=0.116\,ns, assuming that the Ir/Pt interface transparency is $t$=0.105.
We note that the curve is not a great fit to the data, the $\chi^2$ is 27.8 with 6 degrees of freedom.
The discrepancy between the data and the model is mainly at small Pt film thicknesses.
A plausible cause could be that \replaced[id=GW]{the approximate solution of the Usadel equation~\cite{Usadel70} has a large error due to the dependence of resistivity on thickness for thin films}{the impurity averaged Usadel Green’s function~\cite{Usadel70} has a large error at a very thin Ir film}.
If we release the constraint of Ir/Pt interface transparency, we find similar fit results: $t$=0.094, $\tau_s$=0.076\,ns, and $\chi^2$=24.3 with 5 degrees of freedom.
\subsection{Fabrication reproducibility}
We carried out preliminary investigations of the stability of our thin film growth process by measuring transition profiles for films grown using different processes. Our results are not comprehensive, but provide a qualitative sense that our room temperature process is reasonably reliable.

\begin{figure}[hbt!]
\begin{center}
\includegraphics[width=3.2 in]{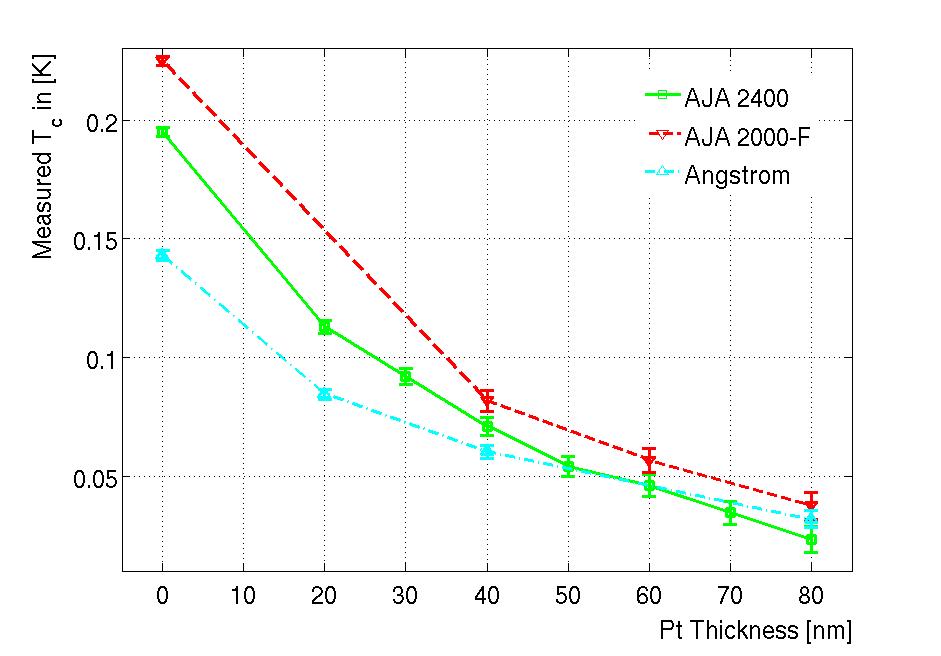}
\includegraphics[width=3.2 in]{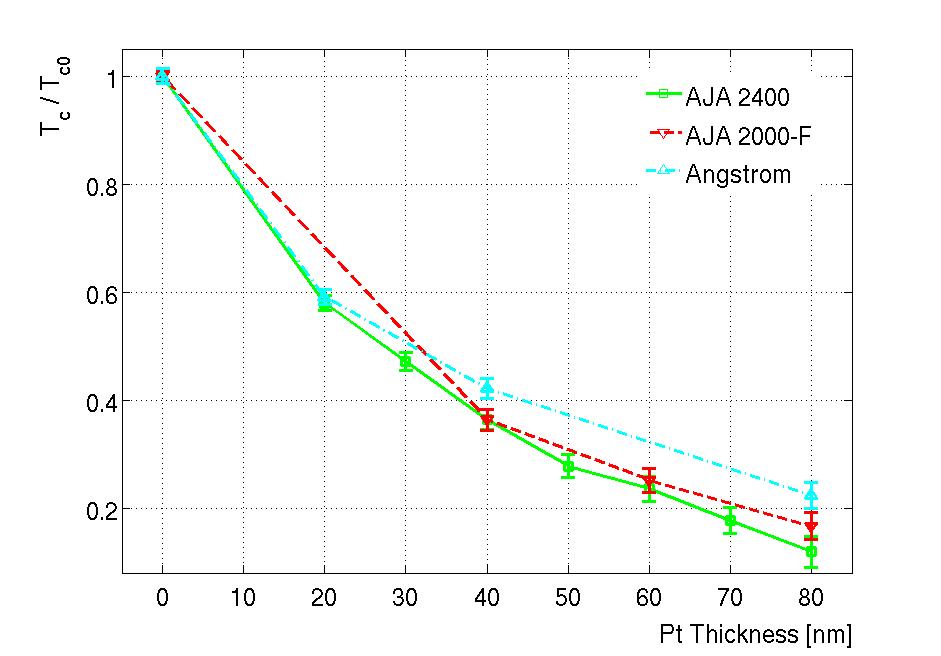}
\caption{
Top: Superconducting critical temperature $T_c$ as a function of Pt film thickness in Ir/Pt bilayers.
The $T_c$ is the measured value with a small Joule heating power correction using the electron-phonon decoupling strength in section~\ref{electron_phonon}.
The thicknesses of Ir films are 100~nm.
The error bars include the $T_{c}$ uncertainties from measurements and the $T_{c}$ uncertainties interpolated from the 3\% film thickness uncertainties.
The bilayers were grown in three different sputtering systems at nominal conditions (AJA 2400: green squares, AJA 2000-F: inverted red inverted triangles, and Angstrom: cyan triangles).
Bottom: The reduced transition temperature ($T_c / T_{c0}$) vs the thickness of Pt films.
}
\label{fig:IrPt_All}
\end{center}
\vspace{-20pt}
\end{figure}

Our first set of studies involved growing Ir/Pt bilayers on two-inch high resistivity silicon wafers using different sputtering systems: in addition to the above devices fabricated on AJA 2400 system, we utilized an AJA 2000-F sputtering system and an Angstrom Engineering sputtering system\added[id=GW]{, both of which are configured for con-focal deposition with tilt sputter sources and provide sample rotation during deposition. The AJA 2000-F system uses a turbo pump for high vacuum.  The Angstrom Engineering sputtering system uses a cryopump for a high vacuum}.
Films grown in the AJA 2000-F used the same Ir and Pt targets as for the films grown in the AJA 2400.
The Angstrom-sputtered films used a three-inch 3N Ir target and a three-inch 4N Pt target.
For these samples made in different sputtering systems, all controllable deposition parameters were the same as those used for the results presented above.
Measured $T_c$ for all of the Ir/Pt bilayers are shown in Fig.~\ref{fig:IrPt_All}.
Our data suggests that our room temperature film growth is fairly reliable with the primary variation appearing to come from changes in the $T_c$ of bare Ir films.

\begin{figure}[ht!]
\begin{center}
\includegraphics[width=3.2 in]{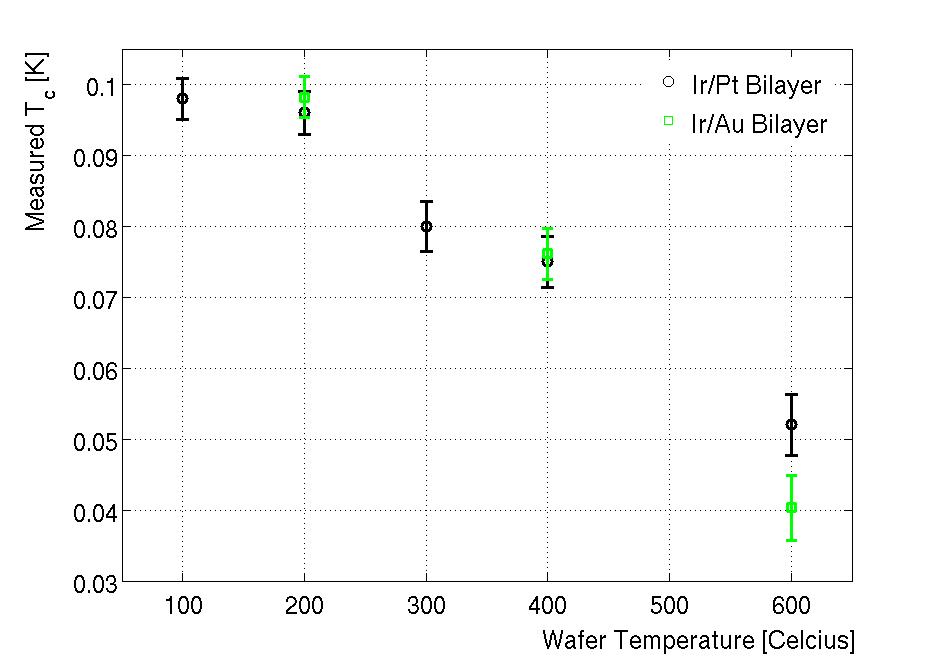}
\caption{
Superconducting critical temperature $T_c$ of Ir/Au bilayers (green squares) and Ir/Pt bilayers (black circles) as a function of growth temperature of 80\,nm thick Ir layer.
The error bars include the uncertainties of temperature measurements and the uncertainties interpolated from a 3\% film thickness controlling uncertainty.
}
\label{fig:IrPt_IrAu_vs_IrTemp}
\end{center}
\vspace{-20pt}
\end{figure}

We also investigated the dependence of the $T_c$ of an Ir/Au and Ir/Pt bilayers on silicon substrate temperature during deposition of the Ir base film.
Similar to the fabrication method in the reference~\cite{CRESST-films-2}, the silicon substrate was heated to an elevated temperature during the Ir film deposition.
Once the Ir deposition was complete, the sample was cooled to room temperature at which point we proceeded to sputter Au or Pt film.

Superconducting critical temperature dependence of Ir/Au bilayers (Ir 80~nm/Au 160~nm) on substrate growth temperature during the Ir film deposition is shown in Fig.~\ref{fig:IrPt_IrAu_vs_IrTemp}.
We see additional $T_c$ suppression at high growth temperatures: the critical temperature of a bilayer grown at 600$^{\circ}$C is 50~mK lower than that of a bilayer grown at 200$^{\circ}$C.
Similar trend is seen in Ir/Pt bilayers (Ir 80\,nm/Pt 20\,nm, black circles in Fig.~\ref{fig:IrPt_IrAu_vs_IrTemp}).
These results are consistent with previous work~\cite{CRESST-films-2}.
The additional reduction of $T_c$ by applying heat during deposition of the Ir base layer could be explained by improved  crystalline structure of the Ir films~\cite{Gong-APS}.
\section{Conclusion} \label{Sec:Conclusion}
We have developed two low $T_c$ film fabrication recipes using sputtering deposition at room temperature.
One is a Au/Ir/Au trilayer, which has a tunable $T_c$ down to 20 mK and features with a low normal resistance.
Compared to an Ir/Au bilayer, a Au/Ir/Au trilayer suppresses the superconducting order parameter from both sides therefore more effectively.
The other is an Ir/Pt bilayer, which has a tunable $T_c$ down to 20 mK.
The normal metal Pt suppresses $T_c$ more effectively than Au allowing for thinner films with less heat capacity and more flexibility in defining TES operational resistance.
Our results suggest that the enhanced $T_c$ suppression by the Pt can be explained by the combination of a large electron density of states at Fermi energy together with the paramagnetic properties of Pt.
We studied films fabricated using different sputtering systems under the same growth conditions and find that our Ir/Pt films are reasonably reliable and reproducible.
Our studies of films using a heated substrate show that additional $T_c$ suppression is possible through substrate heating.
However, there are immediate benefits in making a low $T_c$ film at room temperature, such as using lift-off patterning without baking photoresist to a temperature above 500 $^0$C, and making a TES directly on the bulk of dielectric crystals which can be very sensitive to heating. \added[id=GW]{Additionally, the room temperature growth may allow for TES fabrication to take place near the end of detector processing eliminating any potential impact from other fabrication steps.}
This work is \replaced[id=GW]{important}{relevant} for \added[id=GW]{high precision} TES calorimeter applications\replaced[id=GW]{, which include}{such as} neutrinoless double beta decay searches\added[id=GW]{, coherent elastic neutrino nucleus scattering measurements~\cite{formaggio2012, billard2017}}, and\added[id=GW]{ low mass} dark matter particle \replaced[id=GW]{detection}{searches} \added[id=GW]{using superfluid helium~\cite{schutz2016, hertel2019, guo2013} or dielectric crystals~\cite{knapen2018, trickle2020, griffin2020}}.
\begin{acknowledgments}
We would like to thank Paul Barton and Jeff Beeman for help dicing the samples and J.G. Wallig for engineering support.
This work was supported by the US Department of Energy (DOE) Office of Science, Office of Basic Energy Sciences under Contract Nos. DE-AC02-05CH11231 and DE-AC02-06CH11357, by the DOE Office of Science, Office of Nuclear Physics under Contract No. DE-FG02-08ER41551, and by the National Science Foundation under grants PHY-0902171 and PHY-1314881.
\end{acknowledgments}
\vspace{7mm}
\noindent \textbf{DATA AVAILABILITY}
\par
\vspace{4mm}
The data that support the findings of this study are available from the corresponding author upon reasonable request.
\nocite{*}
\bibliography{aipsamptc}
\end{document}